\documentclass[a4paper]{jpconf}
\usepackage{graphicx}
\usepackage{amssymb}
\usepackage{amsmath}
\usepackage[square,sort&compress]{natbib}
\begin{document}
\title{High-resolution radio observations of X-ray binaries}

\author{J C A Miller-Jones$^{1,}$\footnote[2]{Jansky Fellow, NRAO.}}

\address{$^1$ NRAO Headquarters, 520 Edgemont Road, Charlottesville, VA,
  22903, USA}

\ead{jmiller@nrao.edu}

\begin{abstract}
I present an overview of important results obtained using
high-resolution very long baseline interferometry (VLBI) observations
of X-ray binary systems.  These results derive from both astrometric
observations and resolved imaging of sources, from black holes to
neutron star and even white dwarf systems.  I outline a number of
upcoming developments in instrumentation, both new facilities and ongoing
upgrades to existing VLBI instruments, and I conclude by identifying a
number of important areas of investigation where VLBI will be crucial
in advancing our understanding of X-ray binaries.
\end{abstract}

\section{Introduction}

Very Long Baseline Interferometry (VLBI) has provided us with some of
the highest resolution observations to date of radio-emitting systems
throughout the visible Universe.  X-ray binary systems, accreting
compact objects in orbit with less-evolved donor stars, emit
synchrotron radiation from their relativistic jets.  During outbursts,
the radio emission from these sources becomes bright, reaching levels
of several Jy in the brightest sources, and is often resolved with
high-resolution observations, making such sources ideal targets for
study with VLBI arrays.

Being Galactic systems, X-ray binaries are much closer than their
scaled-up extragalactic analogues, the Active Galactic Nuclei (AGN).
However, since they are typically $10^6$--$10^8$ times less massive
than AGN, but only $10^4$ times closer (a few kpc compared to several
tens to hundreds of Mpc), it is, counterintuitively, possible to probe
closer to the compact object (in units of gravitational radii) in AGN
than in X-ray binary systems, so the Galactic systems are in fact less
useful for studying the close-in regions where the jets are
accelerated and collimated.  Furthermore, since the number of X-ray
binaries with resolved radio emission is still small, studies of these
sources are to some extent limited by the peculiarities of individual
objects.

However, studies of X-ray binary systems do have some advantages over
AGN.  Their proximity means that fundamental parameters such as
distance and space velocity can be determined by astrometric
measurements.  And since lengths and timescales close to the compact
object scale with black hole mass, it is possible to observe the
evolution of the jets in X-ray binaries on timescales of hours to
days, rather than years to decades, following the evolution of the
source through its entire duty cycle in a matter of months, and
watching the evolution of the radio jets in real time as they move out
and interact with their environments.  Furthermore, since the class of
X-ray binaries comprises accreting neutron stars as well as black hole
systems, they allow us to probe the importance of a deep potential
well and the existence of a solid surface and a stellar magnetic field
in the acceleration and collimation of relativistic jets.

In this article I will attempt to provide a brief overview of recent
high-resolution studies of X-ray binary systems.  In this definition,
I choose to include all accreting Galactic compact objects, from black
holes through neutron stars, and even white dwarf systems.  I will
outline new and upcoming developments in instrumentation, as well as
their potential applications in solving some of the important open
questions in the field of X-ray binary research.

\section{Astrometry}

\subsection{Distances via trigonometric parallax}

Distances to X-ray binary systems are notoriously poorly-constrained
and model-dependent \citep{Jon04}, despite being crucial to the
interpretation of almost any astronomical measurement.  Accurate
distances are necessary for converting fluxes into luminosities,
angular sizes into physical separations, and proper motions into
speeds.  A trigonometric parallax is the only direct,
model-independent method of distance determination available, although
at typical X-ray binary distances of a few kpc, the expected parallax
signal is less than one milliarcsecond.  Therefore, high-resolution
astrometric VLBI observations can, in principle, provide a parallactic
distance to relatively nearby Galactic sources.  Unfortunately, black
hole X-ray binary systems are not particularly well-suited to such
measurements.  They spend the majority of their lives in a quiescent
state, only becoming sufficiently bright to be detected in the radio
band in a reasonable integration time during occasional outbursts.
During such outbursts, the sources exhibit relativistically-moving
radio jets, making it difficult to accurately determine the position
of the binary itself, adding scatter to any measurement of the
parallax.  Furthermore, occurring on an unpredictable and irregular
basis, these outbursts are typically not sufficiently well-timed
(i.e., not occurring at three- or six-month intervals) for a
parallactic distance determination.  For this reason, only two
reliable parallactic distances have been measured to date, in the
systems Sco X-1 \citep{Bra99} and Cyg X-1 \citep{Les99}.

Other methods of distance determination are subject to a variety of
systematic errors, which were found to artificially enlarge the
difference between the quiescent X-ray luminosities of black hole and
neutron star systems \citep{Jon04}.  If so, this would undermine the
claimed evidence for black hole event horizons \citep{Men99,Gar01}.
Accurate distance determinations for black hole X-ray binaries would
allow better estimations of their peak luminosities, thereby
constraining the factor by which they can exceed their Eddington
luminosities, which has important implications for whether the
Ultraluminous X-ray sources (ULXs) in external galaxies are
intermediate-mass black holes, or simply super-Eddington stellar-mass
X-ray binaries.  Constraining the systematics affecting other methods
via model-independent parallax distance measurements to a handful of
X-ray binaries would help tie down the distance scale for the majority
of systems, allowing us to statistically verify the peak and quiescent
luminosity distributions.

\subsection{Kinematics via proper motion measurements}

A second use of high-precision astrometry is to measure the proper
motion of X-ray binary systems.  Together with a radial velocity
measurement, the Galactic co-ordinates and the distance of the source,
a measured proper motion allows the full three-dimensional space
velocity of the system to be calculated.  Given other fundamental
system parameters such as the constituent masses, and the luminosity
and temperature of the donor star, the full evolutionary history of
the binary can be reconstructed back to the time of compact object
formation, as has been done for GRO J\,1655-40 \citep{Wil05}.

Since the majority of X-ray binaries are at kiloparsec distances,
their proper motions are of order a few milliarcseconds per year, so
high-angular resolution VLBI measurements are typically required.
This has so far been done for only a handful of X-ray binary systems
\citep{Mir01,Mir02,Rib02,Mir03a,Dha07}, although optical observations
(e.g., with the {\it Hubble Space Telescope}), occasionally in concert
with radio data, have also yielded proper motions
\citep{Mir03b,Mig02,Gon05}.  The measured proper motions can be
compared with those expected from Galactic rotation, and the
discrepancy (the peculiar motion) gives us information about the
formation mechanism of the binary.  Systems with a high velocity out
of the plane or in an eccentric orbit about the Galactic Centre, such
as Sco~X-1 \citep{Mir03b}, XTE~J\,1118+480 \citep{Mir01} or LS\,5039
\citep{Rib02} must either have been born in the halo, or received a
substantial natal kick during their formation in a supernova.  On the
other hand, the most massive black holes ($>10M_{\odot}$) are believed
to form without an energetic supernova explosion \citep{Fry01}, as
demonstrated by the case of Cygnus X-1, where $<1M_{\odot}$ is
believed to have been ejected in the natal supernova \citep{Mir03a}.
More statistics on black hole X-ray binary space velocities (from
accurate proper motions, distances and black hole mass estimates) are
needed to confirm or refute this scenario.

Without accurate proper motions, the radio morphologies exhibited by
X-ray binaries can be difficult to interpret.  For example, in the
case of Cygnus X-3, a bright, well-studied radio-emitting X-ray
binary, the morphology of the observed radio jets seems to change from
epoch to epoch, likely owing to varying interactions with the dense,
inhomogeneous circumstellar medium created by the Wolf-Rayet wind of
the donor star.  The source has been seen to show both one-sided
\citep{Mio01} and two-sided \citep{Mil04} jets, with varying
orientations and degrees of curvature of the jets at different epochs.
An accurate core position for any given epoch (which can be obtained
from a proper motion) will aid the interpretation of some of the more
complicated VLBI structures seen from the jets \citep[e.g.,][]{Tud07}.

\subsection{Resolving the binary orbit}
\label{sec:lsi}
Precision astrometry can even, in some cases, resolve the orbit of the
compact object in a binary system, as done for the case of
LSI$+61^{\circ}303$ in a beautiful demonstration of the power of
high-resolution VLBI techniques.  This source is a high-mass X-ray
binary system which has been observed at all wavelengths from the
radio band up to gamma-ray and even TeV emission.  It comprises a
1--3\,$M_{\odot}$ accretor in an elliptical 26.5-d orbit (with a
semi-major axis of $\sim0.4$\,mas) with a $12M_{\odot}$ donor star.
Resolved radio emission has been observed from the source, originally
interpreted as a precessing, relativistic jet \citep{Mas01,Mas04}.
With Very Long Baseline Array (VLBA) observations at 10 epochs over
the 26.5-d orbit, Dhawan et al.\ \citep{Dha06} tracked the radio
source over the course of the orbit.  They found the radio emission to
have a cometary morphology, with the axis of the emission pointing
away from the high-mass donor star at all times, and varying
dramatically at periastron.  From this, they deduced that the radio
emission arose not from a jet, but from electrons shock-accelerated in
a pulsar wind where it interacts with the dense wind of the donor
star.  This proves that the source is a Be-star/pulsar binary, rather
than a typical microquasar, which would be expected to have a resolved
radio jet.

\section{Imaging}

Since X-ray binaries evolve on human timescales, we can observe their
morphology change in real time.  From the proper motions of components,
we can constrain the jet velocity and orientation, probe the details
of any interactions between the jets and their environment, and
observe jet precession in the sources where it occurs.  Furthermore,
observations which can resolve the jets perpendicular to their
direction of motion can, under the assumption of no confinement, place
limits on the Lorentz factor of the jet flow \citep{Mil06}.

\subsection{Outflow morphology}

High-resolution imaging is the only method to definitively ascertain
the morphology of the radio emission from X-ray binaries.  While jet
synchrotron radiation, either from steady conical outflows, or from
transient discrete ejecta, is the default interpretation of most
unresolved radio emission, recent work has demonstrated that not all
radio emission from X-ray binaries takes the form of collimated
relativistic jets.  LSI$+61^{\circ}303$ (Section \ref{sec:lsi}) shows
a cometary morphology due to interactions between a pulsar wind and
the stellar wind of the donor, and high-resolution studies of the B[e]
X-ray binary CI Cam \citep{Mio04} showed an ellipsoidal or double-ring
morphology which expanded over the course of a year.  This was
interpreted as the propagation of a shock wave formed as the inner
jets interacted with the dense surrounding circumstellar medium
produced by the strong stellar wind of the donor star.  Nevertheless,
many systems do show jet-like emission, of which one of the most
well-studied is SS\,433.

VLBI observations have significantly enhanced our understanding of the
X-ray binary system SS\,433.  This source, accreting at
super-Eddington rates, exhibits precessing, antiparallel jets moving
at a speed of $0.26c$.  The jets are thought to have inflated the
W\,50 nebula surrounding the source.  The precession of the jets is
well described by the kinematic model \citep{Mil79,Abe79,Fab79}, and
has been beautifully verified by VLBI imaging.  Not only does the
locus of radio emission follow the model predictions, but proper
motion measurements on individual components give speeds which agree
with the velocity derived from fits to the kinematic model.
High-resolution imaging has also provided evidence for a wealth of new
and unexpected phenomena in the source, including the existence of a
brightening zone at $\sim250$\,AU from the core \citep{Ver87}, and the
presence of an equatorial outflow, possibly a wind from the accretion
disc \citep{Par99,Blu01}.

In the most intensive VLBI monitoring campaign on an X-ray binary to
date, SS\,433 was observed with the VLBA at 1.5-GHz over one full
quarter of its 162.5-d precession period\footnote[1]{\texttt
http://www.aoc.nrao.edu/$\sim$mrupen/XRT/SS433/ss433.shtml}.  The
proper motion of the equatorial emission was detected, implying a
velocity of $\sim10,000$\,km\,s$^{-1}$, and an azimuthal dependence of
the brightening region was observed \citep{Mio04b}.  A full analysis
of this rich dataset is yet to be published.

\subsection{Canonical black hole jets}

High-resolution observations of black hole X-ray binaries have shown
that the jets exist in two distinct forms.  The most spectacular jets
are the bright, optically-thin, relativistically-moving knots of
emission seen during transient outbursts
\citep[e.g.,][]{Mir94,Tin95,Hje95,Fen99}.  Such knots have always been
seen to move ballistically away from the core, fading as they move
outwards \citep{Mil07}.

In the quiescent state, where the systems spend most of their time,
the jets take the form of steady, partially self-absorbed, conical
outflows.  In this state, the jet length scales as $\nu^{-1}$, since
we see the $\tau=1$ surface at any frequency, which is closer to the
core at higher frequencies.  Such jets have only been directly imaged
in the two systems, GRS\,1915+105 \citep{Dha00} and Cygnus X-1
\citep{Sti01}, but are inferred to exist in all hard-state and
quiescent systems, due to the observed flat radio spectra and the
unbroken correlation between X-ray and radio emission seen in hard and
quiescent systems \citep{Gal03}, extending from a few per cent of the
Eddington luminosity ($L_{\rm Edd}$), where the jets have been directly
imaged \citep{Dha00,Sti01}, all the way down to $10^{-8.5}L_{\rm
Edd}$.  To date, no jets have ever been resolved in low-luminosity
($<0.01 L_{\rm Edd}$) systems.

A unified picture of the jet-disc coupling puts these two types of jet
into context, associating them with particular types of X-ray emission
over the entire duty cycle of a black hole X-ray binary \citep{Fen04}.
As a black hole system emerges out of quiescence, the power in the
steady, partially self-absorbed, conical jets increases with the X-ray
intensity, although the X-ray spectrum remains hard.  The X-ray
spectrum eventually begins to soften, and the jet velocity increases,
causing internal shocks to appear within the flow as the fast-moving
jets collide with slower material further downstream, which are seen
as bright, transient, relativistically-moving ejecta.  The radio core
then switches off, and the source shows a soft, blackbody X-ray
spectrum from the accretion disc.  There may be several transitions
between the soft and hard states at high X-ray intensity, each
corresponding to a new transient ejection event, before the source
fades, the X-ray spectrum gets harder, the core jet is re-established,
and the system moves back into quiescence.  However, the radio aspect
of this picture still remains to be observationally verified, since
the jet morphology, of both the steady and transient jets, has only
ever been resolved during the highest-luminosity part of the duty
cycle.

\subsection{Neutron star jets}

While jets have been relatively well-studied in black-hole X-ray
binary systems, their neutron-star counterparts remain much more
enigmatic.  The intrinsic faintness of the jets in neutron star
systems, which are $\sim30$ times fainter in the radio band at a given
X-ray luminosity \citep{Mig06}, means that current instrumentation has
the sensitivity to detect only the very brightest systems.  These are
the Z-sources, a subclass of the low magnetic field neutron star
systems which are persistently accreting at or near the Eddington
rate.  A model for the jet-disc coupling has been proposed for neutron
stars \citep{Mig06}, analogous to that developed for the black hole
systems, which suggests that steady jets exist in the hardest X-ray
states, which evolve to transient ejecta at transitions to softer
states, when the nuclear jet switches off.  Again, this picture
remains to be verified by direct, high-resolution imaging.

The most detailed high-resolution study of a neutron star X-ray binary
system to date was a set of monitoring observations of Sco X-1
\citep{Fom01}.  The source was seen to be dominated by the emission
from the working surfaces where a highly relativistic beam of plasma
from the core impinged on the ISM.  These lobes were stable in
position angle over several years, although different lobe advance
speeds were seen at different times.  Intriguingly, the other
confirmed neutron star system in which jets have been resolved, Cir
X-1, also showed evidence for a highly-relativistic unseen outflow
illuminating downstream radio lobes \citep{Fen04}, and has the highest
Lorentz factor observed to date from an X-ray binary system ($v_{\rm
app}=9.2c$ \citep{Iar05}).  This strongly argues against the theory
that jet velocity should scale with the escape speed from the compact
object.

\subsection{White dwarf jets}

Recent observations have demonstrated that jets are not only confined
to black hole and neutron star systems.  The recurrent nova RS Oph
consists of a white dwarf accreting from the wind of its red giant
companion.  The white dwarf is close to the Chandrasekhar mass, and
every $\sim 20$\,y, a thermonuclear runaway occurs on its surface,
burning the accreted layer of hydrogen.  The ejected material from
this explosion interacts with the wind of the red giant donor, and the
resulting shocks generate radio emission.  MERLIN and VLBA images of
the 2006 outburst \citep{OBr06,Rup07} showed an expanding spherical
source, thought to be the shock where the explosion was propagating
through the red giant wind, together with an additional collimated
component moving at close to the white dwarf escape speed of
$\sim17,000$\,km\,s$^{-1}$.  This is a clear demonstration of the
presence of jets in a white dwarf system.  Jets have previously been
imaged in the symbiotic star CH Cyg \citep{Tay86} and recent work
shows strong indications of the presence a jet during an outburst of
the dwarf nova SS Cyg \citep{Koe08}, suggesting that high-resolution
imaging of the many different classes of white dwarf systems could
produce interesting new insights into such sources.

\section{Polarization}
Measurements of the polarization of the jets in X-ray binaries can
give us information about the magnetic fields and how they align with
the jet axis.  As an example, full polarization imaging has been used
in AGN jets to reconstruct the emissivity, velocity and magnetic field
distributions in the jets \citep{Lai02}.  Synchrotron radiation theory
predicts a maximum fractional polarization of order 70 per cent for
optically thin radiation.  Observed degrees of polarization in radio
observations of X-ray binaries (with typical resolutions of a few
arcseconds), where it is detected at all, are typically of order a few
to 20 per cent \citep[e.g.,][]{Hje81,Han92,Han00,Fen02,Bro07}.  Such a
discrepancy with the theoretical maximum implies that there must be
some line-of-sight, beam, or external Faraday depolarization
occurring.  With higher-resolution (i.e., VLBI) observations, we can
hope to reduce the effect of beam depolarization, and obtain
information about the true magnetic field structure.
Full-polarization VLBI imaging of X-ray binaries has only been carried
out in a limited number of cases to date.  MERLIN observations found
the central regions of SS\,433 \citep{Par99b,Sti04} to be depolarized,
whereas GRS\,1915+105 \citep{Fen99,Mil05} and Cygnus X-3 \citep{Tud07}
showed between a few and 25 per cent polarization in the approaching
jets.  To have a hope of disentangling the magnetic field structure in
X-ray binary jets, more high-resolution full-polarization observations
are needed at high frequencies, where Faraday depolarization is less
important.

\section{New developments in instrumentation}

\subsection{e-VLBI}
Recent years have seen the advent of eVLBI, very long baseline
interferometry conducted in real time, transporting the data over the
internet to a central location for correlation.  This has made
rapid-response science possible.  Instead of waiting for a few weeks
for the data to be shipped to the correlator, correlation is performed
in real time, and the data can be reduced immediately.  For sources
such as X-ray binaries which evolve on a timescale of hours to days,
this allows decisions on future observing strategies to be refined in
the light of current data.  As the source evolves from optically thick
to thin, and expands, the observing frequency can be changed to
provide optimal spatial sampling of the emitting regions.
Decisions on the appropriate time sampling can be made according to
the rate at which the source evolves.  Should the source be
scatter-broadened, or the phase reference calibrator be found to be
too faint, adjustments to the observing strategy can be made.  For
long-term monitoring campaigns, this prevents telescope time from
being used at times after the source has faded below detectability.

Early results of eVLBI imaging from both European \citep{Tud07,Rus07}
and Southern Hemisphere \citep{Phi07} arrays have now been published.
The sustainable bit rate for the European VLBI Network (EVN)
observations was only 128\,Mbps, although this has now been upgraded
to 512\,Mbps, and the number of connected stations has increased from
six to eight, providing enhanced {\it uv}-coverage and
sensitivity\footnote[2]{\texttt
http://www.evlbi.org/evlbi/e-vlbi\_status.html}.  By way of
comparison, the eVLBI observations with the Long Baseline Array (LBA)
used a 1\,Gbps data rate between the three mainland stations, with a
reduced bandwidth link to the dish in Hobart\footnote[3]{\texttt
http://www.atnf.csiro.au/vlbi/documentation/VLBI\_National\_Facility\_upgrade.html}.

\subsection{Upgrade to the VLBA}
The ongoing upgrade to the VLBA\footnote[4]{\texttt
http://www.vlba.nrao.edu/memos/sensi/} aims to increase the peak and
standard sensitivities by a factor of 2.8 and 5.6 respectively, by
increasing the data rate from 128\,Mbps to 4\,Gbps by 2011.  This
would provide an image noise level of $<10$\,$\mu$Jy\,beam$^{-1}$ in
8\,h, and an astrometric accuracy of 10\,$\mu$as on a 1-mJy source.
With a comparable noise level, faster slew times, much greater ease of
scheduling, better phase stability, and a larger field of view
(permitting in-beam calibrators), this will allow the VLBA to rival or
surpass the High Sensitivity Array (HSA) for astrometric observations.

The low-noise amplifiers for the 22-GHz VLBA receivers have already
been replaced, resulting in a 38\% decrease in the system noise,
equivalent to increasing the integration time by a factor of 2.5.
Further potential upgrades to wider bandwidth or lower noise levels
are also being considered in other frequency bands where modern
electronics can deliver enhanced performance.  A new software
correlator is being developed, which will be easily scaled (by adding
more processors) to deal with increased bandwidth or data rates.

The improved instantaneous sensitivity will provide time-resolved
lightcurves for faint sources, and allow us to detect jet components
in a shorter integration time, such that they have not moved
significantly over the course of an integration.  This will also allow
the use of fainter calibrator sources, such that closer calibrators
become available for the majority of sources, providing better phase
referencing and allowing more accurate astrometry and the detection of
weaker targets.  We will be able to extend the existing studies of
black hole X-ray binaries to the lower-luminosity hard and quiescent
states, investigating how parameters such as the power, length, and
degree of collimation of the jets vary with mass accretion rate.
Furthermore, it will become feasible to study the jets of neutron star
systems (typically a factor 30 fainter in radio luminosity) with VLBI,
allowing us to probe how the effects of strong gravity, a solid
surface and a stellar magnetic field affect the properties of the
jets.

With the improved astrometric capability, parallactic distances will
be feasible for all Galactic X-ray binaries accessible to the VLBA
(10\,$\mu$as would give a $10\sigma$ detection of a parallax at
10\,kpc).  The orbits of long-period systems ($>90$\,h for a system
containing a $10M_{\odot}$ black hole at 10\,kpc) could be resolved,
as was done for LSI$+61^{\circ}303$ \citep{Dha06}.  Positional shifts
between quiescent and flaring states could also be used to determine
the location along the jets responsible for the radio emission.

VLBI observations of Galactic X-ray binary systems at all but the
highest frequencies are often degraded by interstellar scattering.
Electron density fluctuations in the interstellar medium modify the
refractive index of the plasma, giving rise to angular broadening of
sources, temporal broadening of pulsed emission, and scintillation
\citep[e.g.,][]{Ric90,Goo97}.  To mitigate these effects and attain the
theoretical instrumental resolution and astrometric accuracy, it is
necessary to observe at high frequencies.  To this end, the upgrade to
the 22\,GHz receivers will allow us to overcome the scattering and
image at full resolution sources in highly scatter-broadened regions
such as the Galactic Plane, where most X-ray binaries are located.
Since optically-thin synchrotron spectra tend to be steep
($S_{\nu}\propto\nu^{-0.7}$), sources are fainter at higher
frequencies (where the best resolution is available for a given set of
baselines), once again demonstrating the need for high sensitivity at
the highest frequencies.

\subsection{VSOP-2}

The resolution of ground-based VLBI at a given frequency is limited by
the maximum baseline, i.e., the Earth's diameter.  To attain higher
resolution, it is necessary to use orbiting antennas, as done by HALCA
from 1997--2003.  A new space-based VLBI mission, VSOP-2/ASTRO-G is
under development, as detailed by Tsuboi (these proceedings).  The
available frequencies of 8, 22, and 43\,GHz will complement
ground-based receiver bands, with a maximum resolution of 38\,$\mu$as
at 43\,GHz, and a phase-referencing capability to enable observations
of faint sources at the high frequencies.  While optically-thin
emission from X-ray binary jets will necessarily be faint at such high
frequencies, deep integrations using VSOP-2 will provide the
opportunity to make the highest-resolution images ever taken of X-ray
binary jets.  This may be necessary to resolve the low-power jets
inferred to exist in the hard and quiescent states.

\section{Open questions}

In this review, I have outlined the current state of the field, and
recent developments in instrumentation.  I will now close by
highlighting a number of issues which I believe need to be addressed
with high-resolution VLBI observations.

The existence of collimated jets in quiescent black hole X-ray
binaries is commonly inferred from the flat radio spectra, but still
remains to be directly tested by high-resolution radio imaging.  At
low luminosities, the jet is believed to be the dominant power output
channel \citep{Fen03}, and as such will affect the dynamics of the
accretion flow.  Whether the X-ray emission at low luminosities is
dominated by a radiatively-inefficient accretion flow
\citep[e.g.,][]{Nar94} or whether it can be explained by
synchrotron-emitting electrons at the jet base \citep{Mar05} is still
under debate.  Measuring how the jet power, length, and degree of
collimation vary with X-ray luminosity will provide important inputs
to the theoretical models seeking to explain the quiescent state.

While the current paradigm for the jet-disc coupling throughout the
duty cycle of black hole X-ray binary outbursts \citep{Fen04b} has
become widely accepted, questions are beginning to be raised over some
of the underlying premises on which it is based \citep{Gal08}.  While
it was derived from a compilation of X-ray and radio behaviour from
many different sources (mainly spectral information, with a limited
set of high-resolution imaging observations taken during outburst or
at the high-luminosity end of the hard state, when the sources are
brightest), it has never been rigorously tested by high-resolution
radio imaging from the rise out of quiescence all the way through to
the end of the outburst.  Similarly, the analogous ``unified model''
proposed to explain the phenomenology observed in the Z-sources
\citep{Mig06} also needs to be verified via VLBI imaging.  Despite
their intrinsic faintness, the speed at which the Z-sources move
through their duty cycles (switching states on a timescale of hours)
makes such observations less time-intensive than similar studies on
black hole systems.

Quantifying the differences in the jet morphology and behaviour
between black hole and neutron star systems is another crucial issue
which needs to be addressed by high-resolution observations.  This
will help quantify the role of a stellar surface, a magnetic field,
and the depth of the potential well in jet formation, acceleration and
collimation.  The relative faintness of neutron star systems as
compared to black hole X-ray binaries means that such studies will
require the enhanced sensitivities of the new and upgraded instruments
coming online.

The recent work on CI Cam \citep{Mio04}, RS Oph \citep{OBr06,Rup07}
and LSI$+61^{\circ}303$ \citep{Dha06} has shown that radio emission
from X-ray binaries should not be blindly interpreted as a collimated
jet.  Such systems exhibit a wide variety of radio morphologies, and
high-resolution imaging of any new outburst is necessary to ascertain
the geometry of the emission, and provide detailed information on the
physical processes occurring in the source.

Verification of the systematics affecting the distance scale by
measuring trigonometric parallaxes to Galactic radio-emitting X-ray
binary systems will provide more accurate derivations of fundamental
system parameters, allowing us to verify the claimed evidence for
event horizons and casting light on the likely nature of ULXs.

Lastly, proper motion measurements for a larger sample of X-ray
binaries are needed to improve the statistics on natal supernova
kicks, and to help constrain black hole formation mechanisms and
binary evolution scenarios.

\ack 
It is a pleasure to thank all my collaborators, in particular
Walter Brisken, Vivek Dhawan, Rob Fender, Elena Gallo, Simone
Migliari, Amy Mioduszewski, and Michael Rupen.  This work has made use of
NASA's Astrophysics Data System.  The author is a Jansky Fellow of the
National Radio Astronomy Observatory, which is operated by Associated
Universities, Inc., under cooperative agreement with the National
Science Foundation.

\bibliography{millerjones_j_r1.bib}

\end{document}